\documentclass[twocolumn]{aastex631}
\pdfoutput=1
\usepackage{mathtools}
\usepackage{appendix}
\usepackage{bm}
\usepackage{soul}
\usepackage{natbib}
\usepackage{orcidlink}
\usepackage{newtxtext,newtxmath}
\usepackage[T1]{fontenc}
\usepackage{ae,aecompl}
\usepackage{chemformula}
\usepackage{xcolor}

\newcommand{\additions}[1]{#1}
\newcommand{\subtractions}[1]{\iffalse #1 \fi}

\shorttitle{Discovery of Gliese 229Bb}
\shortauthors{Whitebook et al.}
\submitjournal{ApJ Letters}

\begin{document}

\title{Discovery of the Binarity of Gliese 229B, and Constraints on the System's Properties}

\author[0000-0002-6836-181X]{Samuel Whitebook}
\affil{Department of Physics, University of California, Santa Barbara, Santa Barbara, CA 93106, USA}
\affil{Division of Physics, Mathematics, and Astronomy, California Institute of Technology, Pasadena, CA 91125, USA}

\author[0000-0003-2630-8073]{Timothy D.~Brandt}
\affiliation{Space Telescope Science Institute, 3700 San Martin Drive, Baltimore, MD 21218, USA}
\affiliation{Department of Physics, University of California, Santa Barbara, Santa Barbara, CA 93106, USA}

\author[0000-0003-0168-3010]{G. Mirek Brandt}
\affiliation{Department of Physics, University of California, Santa Barbara, Santa Barbara, CA 93106, USA}

\author[0000-0002-0618-5128]{Emily C. Martin}
\affiliation{Department of Astronomy and Astrophysics, University of California, Santa Cruz, Santa Cruz, CA, 95064, USA}
\affiliation{Planet Labs, PBC, San Francisco, CA, 94107, USA}

\correspondingauthor{Samuel Whitebook}
\email{sewhitebook@astro.caltech.edu}

\begin{abstract}

We present two epochs of radial velocities of the first imaged T dwarf Gliese~229~B obtained with Keck/NIRSPEC. The two radial velocities are discrepant with one another, and with the radial velocity of the host star, at $\approx$$11\sigma$ significance.  This points to the existence of a previously postulated, but as-yet undetected, massive companion to Gl~229~B; we denote the two components as Gl~229~Ba and Gl~229~Bb. We compute the joint likelihood of the radial velocities to constrain the period and mass of the secondary companion.  Our radial velocities are consistent with an orbital period between a few days and $\approx$60 days, and a secondary mass of at least $\approx$15\,$M_{\rm Jup}$ and up to nearly half the total system mass of Gl~229~B.  With a significant fraction of the system mass in a faint companion, the strong tension between Gl~229~B's dynamical mass and the predictions of evolutionary models is resolved.

\textbf{Key words:} 
\end{abstract}

\section{Introduction} \label{introduction}

Brown dwarfs lie below the minimum mass needed to achieve core hydrogen fusion \citep{Burrows+Liebert_1993}.  Unable to reach the stellar main sequence, they cool and fade through spectral types M, L, T, and even Y \citep{Kirkpatrick+Reid+Liebert+etal_1999,Kirkpatrick+Gelino+Cushing+etal_2012}.  These cooler spectral types are defined by the prominence of molecules like methane and CO in the atmosphere.  Gl~229~B, discovered in 1995 \citep{Nakajima+Oppenheimer+Kulkarni+etal_1995,Oppenheimer+Kulkarni+Matthews+etal_1995}, was the first unambiguously identified T dwarf; its methane-rich atmosphere was clear evidence that it was substellar.  

Soon after its discovery, the first atmospheric and evolutionary analyses attempted to derive a mass and age for Gl~229~B from its observed spectrum and luminosity \citep{Allard+Hauschildt+Baraffe+Chabrier_1996,Marley+Saumon+Guillot+etal_1996}; these favored masses of $\sim$30--50\,$M_{\rm Jup}$.  A series of observations in the years following Gl~229~B's discovery better established its infrared spectrum and even optical colors \citep{Geballe+Kulkarni+Woodward+Sloan_1996,Golimowski+Burrows+Kulkarni+etal_1998,Oppenheimer+Kulkarni+Matthews+vanKerkwijk_1998,Leggett+Toomey+Geballe+etal_1999,Saumon+Geballe+Leggett+etal_2000}.  Under the first definition of the T spectral class Gl 229B was classified as a T6.5V dwarf \citep{Burrows_2002}, however it was later reclassified to a T7pec due to \ch{CH4} features inconsistent with any standard T dwarf model \citep{Burgasser_2006}.  The volume of literature on Gl~229~B decreased in the years following its discovery as new data did not substantially change the initial atmospheric and evolutionary results.  

Interest in Gl~229~B increased following the derivation of a surprisingly high dynamical mass of $70.4 \pm 4.8 M_\mathrm{Jup}$ \citep{Brandt_2020}, in tension with all substellar evolutionary models.  This dynamical mass relied on absolute astrometry cross-calibrated in the Hipparcos-Gaia Catalog of Accelerations \citep{Brandt_2018,Brandt_2021} together with precision astrometry from the Hubble Space Telescope \citep{Golimowski+Burrows+Kulkarni+etal_1998} and radial velocities (RVs) from the California Planet Survey \citep{Butler+Vogt+Laughlin+etal_2017}.  An updated analysis with better astrometry from the third Gaia data release \citep{Lindegren+Klioner+Hernandez+etal_2021,Brandt_2021} confirmed a mass of $71.4 \pm 0.7\,M_{\rm Jup}$ for Gl~229~B \citep{Brandt+Dupuy+Li+etal_2021}.  At this mass, even 10\,Gyr are insufficient for Gl~229~B to cool to its observed luminosity, while the properties of Gl~229~A suggest an age closer to $\sim$1--3\,Gyr \citep{Brandt_2020}.  \cite{Brandt_2020} and \cite{Brandt+Dupuy+Li+etal_2021} both suggested that binarity could provide a solution: some of Gl~229~B's mass could be held by an unseen companion, with the mass in the primary brown dwarf being consistent with expectations from the earlier evolutionary analyses.  

\begin{figure*}[th!]\centering
    \includegraphics[width =\linewidth]{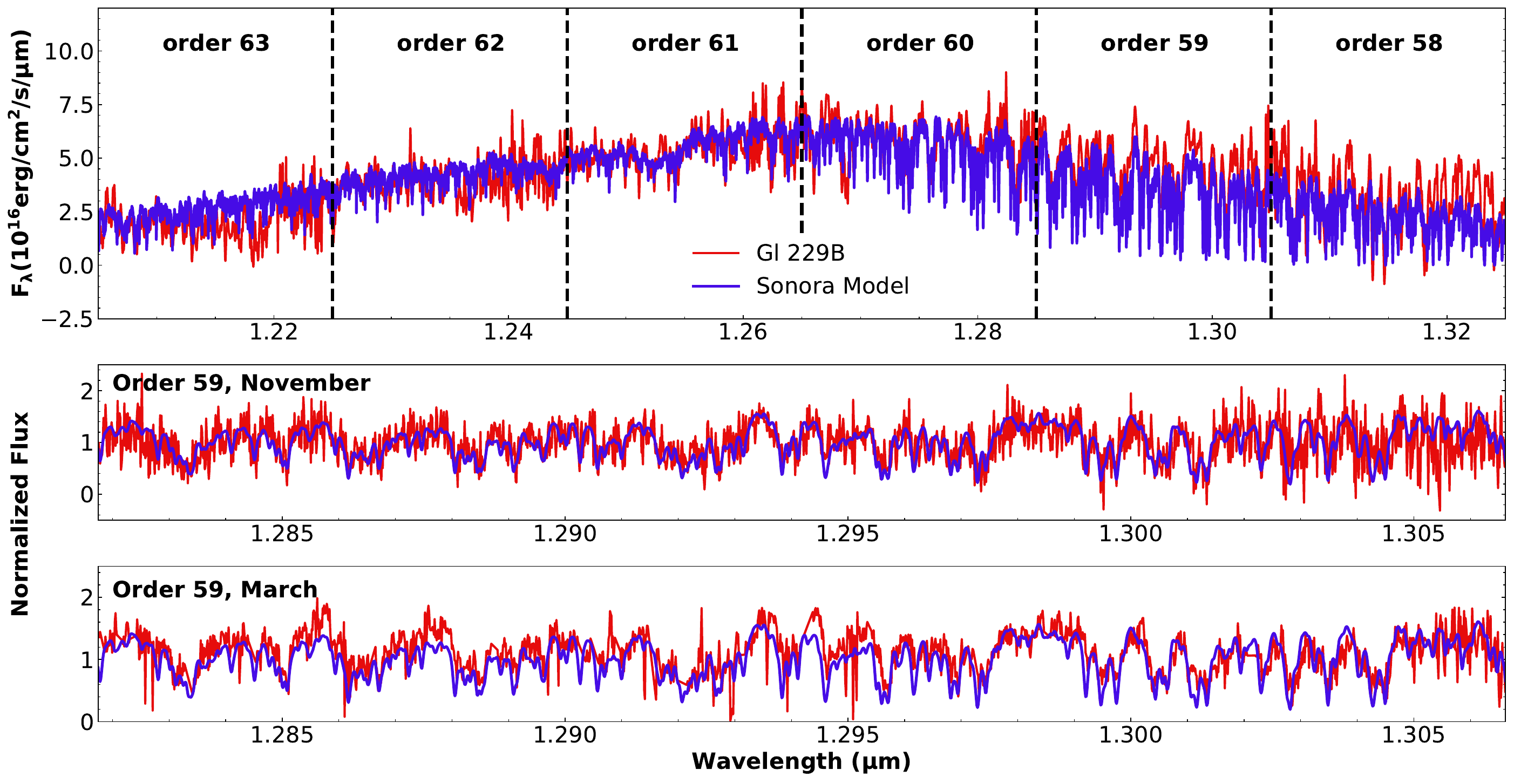}
    \caption{Top: The spectrum of Gliese 229B in the $J$ band between orders 58 and 63 taken in November 2022. The spectrum is flux-calibrated by a sensitivity function derived from Gl 229A \additions{and}\subtractions{The spectrum is} convolved to a resolution of $R \sim 8000$. A \texttt{Sonora} model \citep{Marley_2021} with T = $850$\,K \additions{and} $\log_{10}{(g/{\rm cm\,s^{-2}})} = 5.0$ is overlaid. \additions{Middle}: The highest resolution $R \sim 25000$ spectrum of Gl 229B from order 59 normalized and shifted to the \additions{solar system barycenter rest frame} versus the normalized \texttt{Sonora} model.
    \additions{Bottom: The same full-resolution order 59 spectrum as the middle panel, but for the natural seeing data from March 2022.}}
    \label{fig:dwarf_spec}
\end{figure*}


Following the discovery of a surprisingly high mass for Gl~229~B, a number of retrievals have re-determined its properties, both with \citep{Howe_2023} and without \citep{Calamari+Faherty_Burningham+etal_2022,Howe_2022} the hypothesis of binarity.  If the brown dwarf is single, it points to fundamental problems in substellar modeling.  If Gl~229~B is a multiple, it indicates a complex dynamical structure for the first brown dwarf system to be imaged, and could present new tests for evolutionary models.  Gl~229~Ba and a hypothetical Gl~229~Bb would have the same age and composition, and with precise individual masses, they could enable rare consistency tests of evolutionary models that are possible for a handful of systems \citep{Dupuy+Liu+Leggett+etal_2015,Chen_2022}.  

In this paper we present two epochs of new spectroscopic observations of Gl~229~B to measure its RV.  This can be used to test for additional motion beyond the orbit of the system around its host star, and thereby test for binary, and ultimately measure the individual masses and orbit.  We organize the paper as follows: Section \ref{sec:observations} covers the reduction of spectra taken from KECK/NIRSPEC. In Section \ref{sec:RV_detection} we discuss the measurement of RVs by shifting observations against model spectra. Sections \ref{sec:constraints} and \ref{sec:discussion} discuss parameters derived from the measured RV anomalies and the status of the Gl~229~B system.

\section{Observations and Data Reduction} \label{sec:observations}

\begin{deluxetable}{lccccr}
    \tablewidth{0pt}
    \tablecaption{Observing Log \label{tab:obslog}}
\tablehead{
    \colhead{Date (UT)} &
    \colhead{Target\tablenotemark{a}} &
    \colhead{Band} &
    \colhead{$t_{\rm exp}$ (s)} &
    \colhead{$N_{\rm exp}$} &
    \colhead{$t_{\rm tot}$ (s)}
    }  
\startdata
2022-03-12 & Gl~229~B & $J$ & 300 & 7 & 2100 \\
2022-03-12 & Gl~229~B & $H$ & 300 & 5 & 1500 \\
2022-03-12 & Gl~229~B & $K$ & 300 & 4 & 1200 \\
2022-11-10 & Gl~229~A & $J$ & 15 & 4 & 60 \\
2022-11-10 & Gl~229~B & $J$ & 300 & 20 & 6000 \\
2022-11-10 & Gl~229~A & $H$ & 9 & 4 & 36 \\
2022-11-10 & Gl~229~B & $H$ & 300 & 8 & 2400 \\
2022-11-10 & Gl~229~A & $K$ & 15 & 4 & 60 \\
2022-11-10 & Gl~229~B & $K$ & 300 & 12 & 3600 
\enddata
\tablenotetext{a}{The 2022-03-12 data were taken without adaptive optics; speckles from Gl~229~A are visible in the slit.  The 2022-11-10 data were taken with natural guide star adaptive optics.}
\end{deluxetable}

We obtained spectra of Gl 229B using Keck/NIRSPEC \citep{McLean_1998} on March 12, 2022 and November 10, 2022 UT in high resolution mode ($R \sim 25000$) in the $J$, $H$, and $K$ bands spanning wavelengths from 1.128 - 2.771 $\mu$m. For the March 2022 data, we used NIRSPEC in natural seeing mode with the $0.\!\!''432$ slit, while for the November 2022 data we used natural guide star adaptive optics (NIRSPAO) with the $0.\!\!''041$ slit.  For the latter date we obtained spectra of Gl~229A with short exposure times followed by spectra of Gl~229B with long exposure times.  Table \ref{tab:obslog} lists our observing log. Both epochs used an ABBA nod pattern.

\additions{All} data were reduced using a custom pipeline. We removed amplifier glow and/or a reset anomaly by fitting an exponential to the lowest 100 rows of each column. Spectral orders were cut out by finding the locations of wavelength orders in flat lamps and then rectified by a transformation matrix that linearized the stellar trace along the dispersion direction and minimized the variance of the arc lamp row sums perpendicular to it. In order to avoid the impact of dead and hot pixels, we utilized the Gaussian process regression module $\texttt{astrofix}$ \citep{Zhang_2021} to impute their values. Post-rectification pixels that had more than a 10\% contribution from bad pixels were then masked and weighted to zero during extraction later. Spectra were extracted using the optimal extraction algorithm described in \citet{Horne_1986}; the profiles of the traces were obtained from Gaussian fits to the median of the traces along the dispersion direction. 
Bad pixel masks were created individually for each exposure. Dead and hot pixels were masked from the pre-rectification dark images.  We also masked $>$$3\sigma$ outliers in each exposure, with $\sigma$ robustly estimated from the surrounding $7 \times 7$ pixels by 1.48 times the median absolute deviation. 
Extracted spectra were further cleaned by Fourier transforming the spectra, setting Fourier modes corrupted by fringing from the detector to zero, and then inverting the Fourier transform. These modes are much lower in spatial frequency than the lines used to determine RVs.  The blaze function of each spectrum was found by fitting a 5th degree polynomial to the observed spectrum and subsequently divided out. 

Data taken in March (with natural seeing) had significant diffraction from Gl~229~A resulting in speckle traces overlapping the spectrum of Gl~229~B. This left \additions{two} usable orders \additions{with ${\rm SNR} > 10$}: orders 58 \additions{and 59} in $J$ band. \subtractions{These data were flattened and rectified using the NSDRP pipeline provided by the Keck Observatory.} Traces representing diffraction speckles of Gl~229~A were modeled by Gaussian kernels along the spatial dimension and removed before extracting the spectrum of Gl~229~B. \subtractions{Order 58 lacked arc lamp calibration lines in the bluer half of the spectrum;} We \subtractions{therefore} computed our wavelength solution \additions{in the March data} using OH sky lines. As explained in Section \ref{sec:RV_detection}, we recover the accepted RV of Gl~229~A from diffracted starlight in the March data.

Our November data were taken using natural guide star adaptive optics \additions{(AO)} \citep{Wizinowich+Acton+Shelton+etal_2000} to mitigate the effects of stellar diffraction by enabling the use of a smaller slit. \additions{Because the smaller slit and PSF greatly reduce the relative strength of sky lines,} wavelength solutions \additions{in AO data} were found by fitting Gaussians to observed ArNeXeKr arc lamp lines and then fitting a polynomial transformation between the centers of these Gaussians and catalogued points provided by the Keck Observatory \citep{McLean_1998}.
We calculate the root-mean-square deviation (RMSD) of the residuals between the fitted and tabulated line locations and discard orders with $ \textrm{RMSD} > 0.1$\,\AA\ (including order 58, which we used in the March data). After these initial data cuts we retain 14 orders. We further cut the November data to only consider orders with ${\rm SNR} > 10$, leaving just three orders: order 47 in $H$ band and orders 59 and 60 in $J$ band. Telluric corrections in November were derived directly by dividing the observed spectra of Gl 229A by model spectra. 

\section{Radial Velocity Anomaly Detection} \label{sec:RV_detection}

\begin{figure}
    \includegraphics[width = \linewidth]{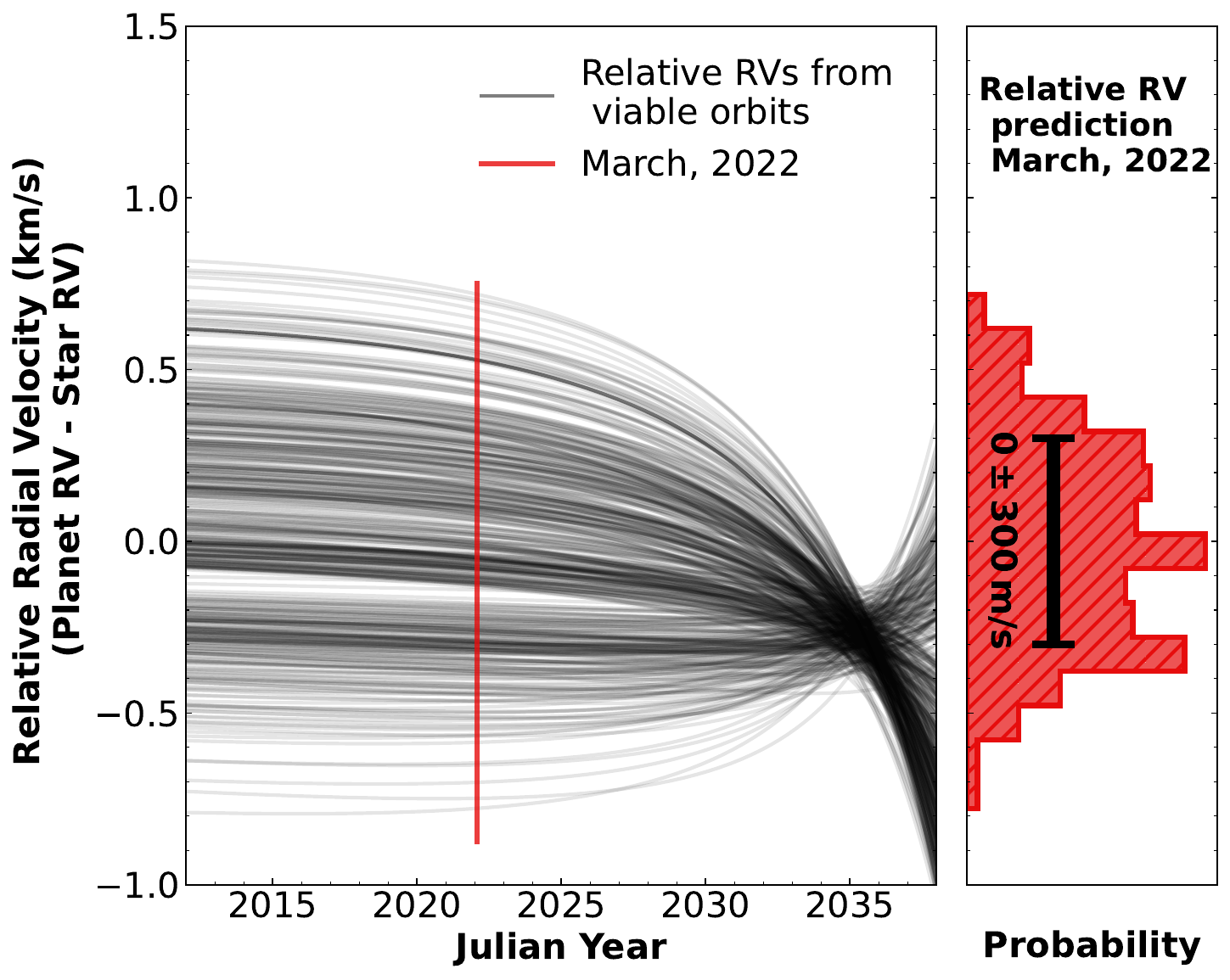}
    \caption{The relative radial velocity of Gl 229B (The RV of Gl 229B minus that of Gl 229A) if the BD is singular.  The orbits are drawn from the posterior in \cite{Brandt+Dupuy+Li+etal_2021}.}
    \label{fig:theoretical_rv}
\end{figure}

Gl 229B has a well-determined orbit \citep{Brandt_2021}. Using the python package \texttt{REBOUND} \citep{Rein_2012}, we predict an RV of $0 \pm 300$\,m\,s$^{-1}$ for Gl 229B relative to Gl~229A in both March and November of 2022 (Figure \ref{fig:theoretical_rv}) where $300$\,m\,s$^{-1}$ is $1 \sigma$ of the \additions{(nearly Gaussian)} distribution of derived orbits. This is the relative RV of the barycenters of Gl~229~A and B, and will be the same as the relative RV of the two objects if Gl~229~B is singular.  \additions{The predicted RV change of a singular Gl 229B across 2022 is very low, $\approx$1\,m\,s$^{-1}$.}

The actual RVs of both Gl 229A and Gl 229B were calculated by finding the redshift that minimizes $\chi^2$ between the observed spectra and relevant rest frame models.
The error used in $\chi^2$ fitting is derived from Poisson statistics. We take the blaze function to be the photon count at every wavelength. The spectra are normalized for fitting against the blaze function, so the variance on the measured spectra is $\sigma^2(\lambda) = 1/F(\lambda)$ where $F(\lambda)$ is the blaze function. 
Additionally, a telluric model derived from \citet{Noll_2012} is used to impart an additional uncertainty added in quadrature to each wavelength bin in correspondence with its absorption.

Gl 229A was modeled using a high resolution M dwarf spectrum generated by the stellar atmosphere code \texttt{PHOENIX} \citep{Husser_2013} with $T_\mathrm{eff} = 3700$\,K and $\log(g) = 4.50$ (cgs) parameters chosen to roughly match Gl 229A. Gl 229B was modeled using the \texttt{Sonora} BD atmosphere models \citep{Marley_2021} with $T_\mathrm{eff} = 850$\,K, $\log{(g)} = 5.0$ (cgs). Each order's model spectrum was normalized by dividing out the best fit first degree polynomial continuum over the wavelength range of the spectral order. This puts the model and the observed spectrum in the same arbitrary unit basis.

Radial velocities calculated by $\chi^2$ were barycenter corrected using the python package \texttt{barycorrpy} \citep{Kanodia_2018}. The November spectra of Gl 229B were taken over the span of approximately 3 hours for $J$ band, and approximately 1 hour for $H$ and $K$ bands, and then averaged; 
spectra of the brown dwarf were barycenter corrected with respect to the mean observation time of each band. We derive uncertainties by bootstrap resampling, \additions{or redrawing data with replacement many times \citep{Efron_Bootstrap},} the individual exposures used to derive RVs. Using \texttt{barycorrpy} we find that the rotation of the Earth over the observational time frame imparts an additional uncertainty of $\approx$$105$\,m\,s$^{-1}$ on the measured RV of Gl~229B; we add this uncertainty in quadrature. Gl~229~A is not subject to this uncertainty, as only a single 30 second exposure of the M dwarf was used.  

Spectra of Gl~229~A were not taken in March. Instead, the RV of the star was derived from diffracted starlight in exposures of Gl~229~B. 
\additions{From the speckle traces of Gl~229~A in the natural seeing data, we derive an RV of $3.3 \pm 1.4$\,km\,s$^{-1}$ for the star in March.  This uncertainty is derived from $\Delta \chi^2 = 1$ and may be less reliable than the one derived using bootstrap resampling of the November data, for which we derive an RV of $3.18 \pm 0.36$\,km\,s$^{-1}$. However, our March RV of Gl~229A agrees well with both our November measurement and with RVs derived in the literature (e.g., 4.7\,km\,s$^{-1}$ from \cite{Nidever+Marcy+Butler+etal_2002} and 4.2\,km\,s$^{-1}$ from Gaia \citep{GaiaDR3}).}
Thus, for the purpose of reporting discrepancies in the theoretical relative RV of the system, we compare the RV of Gl~229~B in both epochs to the RV of Gl~229~A in November.

We report a radial velocity difference of \additions{$-10.35 \pm 0.71$ km\,s$^{-1}$} between Gl 229A and 229B in March 2022, and a difference of $2.74 \pm 1.09$ km\,s$^{-1}$ in November.  These RVs are \additions{$14.5 \sigma$} and $2.5 \sigma$ discrepant, respectively, with the difference of $0 \pm 300\,$m\,s$^{-1}$ expected from the system's known orbit.

\begin{figure}
    \includegraphics[width=\linewidth]{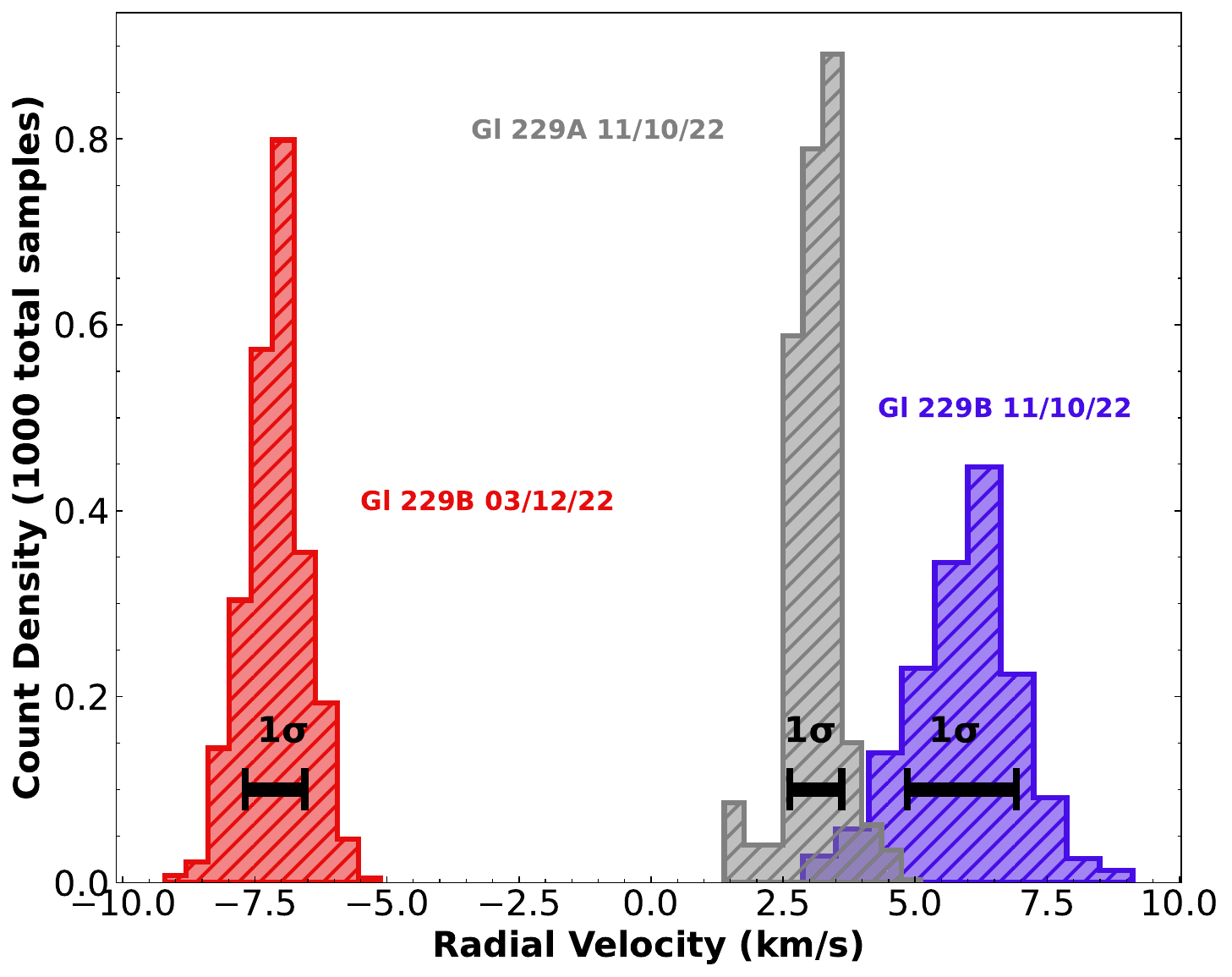}
    \caption{Distributions of individual component RVs in both observation epochs derived from bootstrap resampled spectra. Diffracted spectra of Gl~229~A in March \subtractions{are unsuitable for fitting} \additions{provide a consistent but less precise RV}, so both epochs of Gl~229~B RVs are compared to that of Gl~229~A from November. The RV of the Gl~229~B system varies by more than \additions{$11 \sigma$} between observations, completely ruling out singularity.}
    \label{fig:rv hist}
\end{figure}

\section{Constraints on the Subcompanion} \label{sec:constraints}

Figure \ref{fig:rv hist} shows our two RV measurements of Gl~229~B together with our measured RV of Gl~229~A.  The RV of Gl~229~A is known to be stable at the several m/s level over a period of a year \citep{Tuomi_2014}, orders of magnitude better than our precision.  The March 2022 RV of Gl~229~B differs from the prediction for the barycenter motion of its orbit by nearly \additions{$15\sigma$}, ruling out Gl~229~B's status as a single object.  We now proceed to use our two RVs to constrain the properties of the Gl~229~B system: its orbital period, and the mass of the less massive component, which we denote Gl~229~Bb.  

We measure the likelihood of a mass for Gl~229~Bb and an orbital period of the system assuming a random orbital phase and orientation, and a uniform distribution of eccentricity up to 0.8.  High eccentricities are common in stellar binaries until tidal circularization damps them \citep[e.g.][]{Meibom+Mathieu_2005,Geller+Mathieu+Latham+etal_2021}.  Periods of $\lesssim$10 days are circularized for stars at ages of a few Gyr \citep{Meibom+Mathieu_2005,Geller+Mathieu+Latham+etal_2021}, but the much smaller sizes of brown dwarfs will reduce the efficiency of tidal damping and allow eccentric orbits to remain so down to shorter periods.  
The eclipsing M dwarf-brown dwarf binary TOI-2119, for example, has an eccentricity $\approx$0.3 despite a period of just $\approx$7 days \citep{Canas+Mahadevan+Bender+etal_2022}.  We adopt a lower limit on the orbital period of 1 day, which would correspond to a semimajor axis of about $1.6\,R_\odot$.  We adopt an upper limit of $35\,M_{\rm Jup}$ on the mass of Gl~229~Bb, as the secondary should be less massive than the primary in order to have a lower luminosity.  We do not account for any dependence of the eccentricity on the orbital period; this is a secondary effect on the resulting probability distributions of RVs given the narrow range of periods, from one to a few days, over which tidal circularization could be efficient.  

We use a grid of points in orbital phase and argument of periastron to compute a probability distribution of observing the system at a given fraction of its RV semiamplitude.  These probability distributions depend on eccentricity, while the RV semiamplitude depends on companion mass, period and inclination (the total mass of the Gl~229~B system is accurately known).  We then multiply the probability distributions of RV by the two observational probability distributions, integrate each of these products, and multiply the two results to obtain a likelihood of a given eccentricity, orbital period, and mass of Gl~229~Bb.  Finally, we marginalize these distributions over eccentricity.

\begin{figure}
    \includegraphics[width=\linewidth]{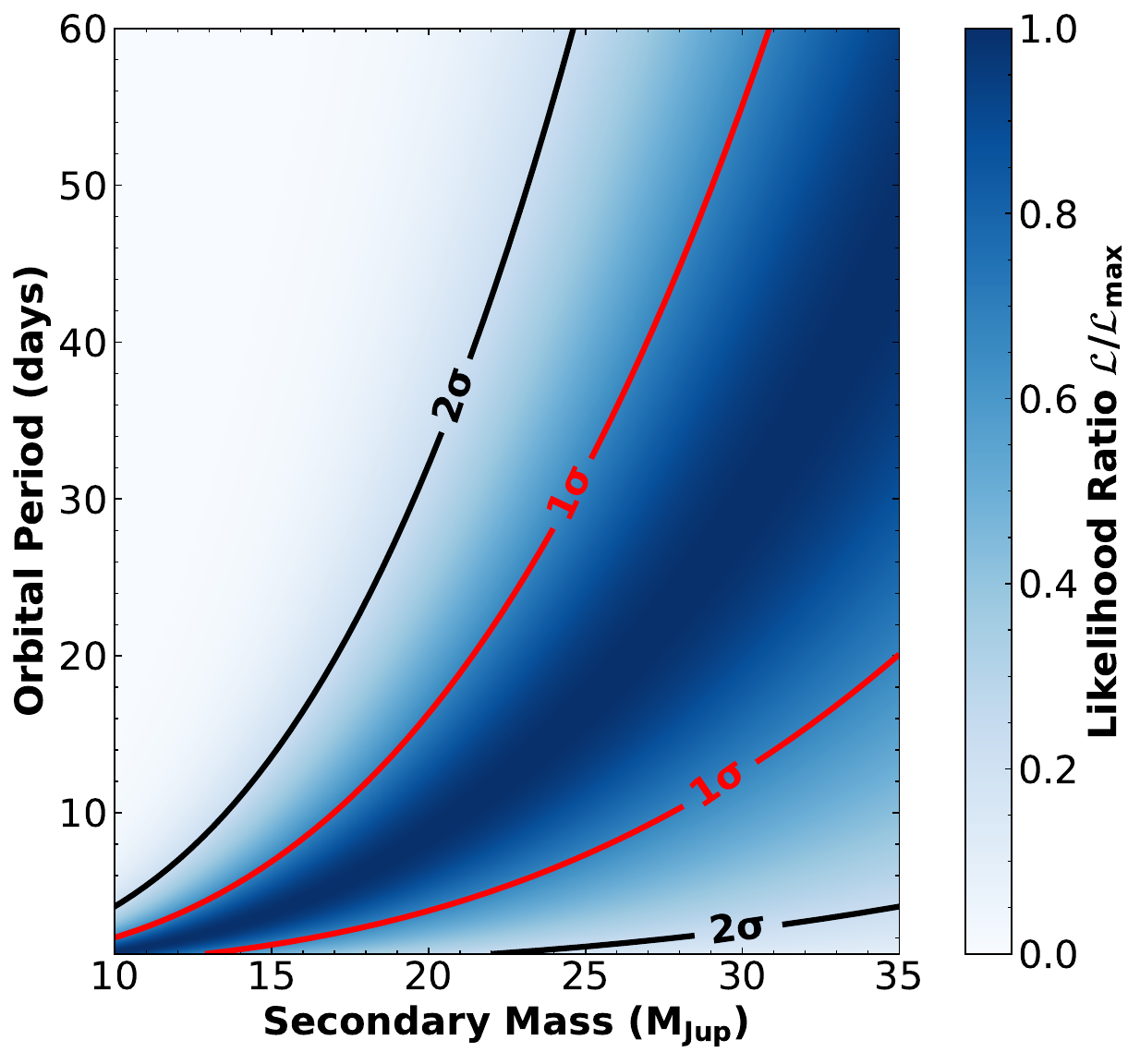}
    \caption{Joint likelihood of orbital period and mass of Gl~229~Bb given our two RV measurements, assuming random eccentricities between 0 and 0.8, random orbital phases and orientations, and a minimum period of 1 day (corresponding to a minimum semimajor axis of $\approx$1.6\,$R_\odot$).  Our two RV measurements confirm that at least $\approx$15\,$M_{\rm Jup}$ of the Gl~229~B system reside in Gl~229~Bb.}
    \label{fig:conditional_probability}
\end{figure}

Figure \ref{fig:conditional_probability} shows our results.  We obtain a mass of Gl~229~Bb of at least $\approx$15\,$M_{\rm Jup}$, with values up to half of the system mass being entirely consistent with the RV data.  Given that Gl~229~Bb should be the less massive body in the system, we constrain the orbital period of the binary to be $\lesssim$\additions{120}\,days at $2\sigma$ and at the highest possible mass of Gl~229~Bb, and $\lesssim$\additions{50}\,days at $1\sigma$ assuming the mass ratio between Gl~229~Ba and Gl~229~Bb to be at least somewhat less than unity.  This constraint establishes Gl~229~Bb as the unseen reservoir for a substantial fraction of Gl~229~B's mass.

Our RV orbital constraints at the longer allowable periods are in mild tension with the results of \cite{Brandt_2020} and \cite{Brandt_2021}.  Those authors disfavored an orbital semimajor axis of Gl~229~Ba of more than a few mas based on the excellent agreement between Hubble Space Telescope relative astrometry and an orbital fit with Gl~229~B as a single object.  This qualitative upper limit on semimajor axis is a few 0.01\,au at the distance to Gl~229, corresponding to an orbital period of a few days assuming a mass ratio close to unity.  The HST observations came in several pairs separated by about six months, so these loose constraints could be evaded depending on the exact orbital period, the alignment of the system, and the possible contribution of Gl~229~Bb to the system's luminosity in the bands observed by Hubble.  However, they do favor periods and masses toward the shorter end of our allowable space, $\sim$20\,$M_{\rm Jup}$ at a period of $\lesssim$1 week.  This would leave Gl~229~Ba as a $\sim$50\,$M_{\rm Jup}$ object, removing the strong tension between the dynamical mass and substellar cooling models \citep{Brandt+Dupuy+Li+etal_2021}.

\additions{Finally, we look for direct spectral evidence of Gl~229~Bb.  At an age of $\lesssim$3\,Gyr \citep{Brandt_2020} and a mass $\gtrsim$15\,$M_{\rm Jup}$, Gl~229~Bb would have an effective temperature of at least $\approx$450\,K \citep{Phillips+Tremblin+Baraffe+etal_2020}, corresponding to a spectral type no later than early Y.  Assuming the same radius for both objects, we fit the sum of two model atmospheres at different temperatures to our observed spectra, but find no pair of temperatures to provide a consistently better fit across spectral orders than a single temperature model.  }

\section{Discussion and Conclusion} \label{sec:discussion}

Models of substellar evolution couple a fully convective interior to an atmosphere with complex chemistry, molecules, and clouds \citep{Burrows+Marley+Hubbard+etal_1997,Allard_2001}.  These models, initially used to infer masses and ages from observed spectra and luminosities \citep[e.g.][]{Allard+Hauschildt+Baraffe+Chabrier_1996,McCaughrean+Close+Scholz+etal_2004}, can now be tested with the aid of dynamical mass measurements and ages from the host stars of substellar companions \citep{Dupuy_2017}.  

\additions{Several late T dwarf companions to main sequence stars have recently been found to have very high masses, close to the stellar/substellar boundary.  WISE J0720-0846B, with a mass of $66\pm 4$\,$M_{\rm Jup}$ \citep{Dupuy+Liu+Best+etal_2019}, can be reconciled with substellar models assuming an old age for the system.  $\varepsilon$~Indi~C, at $70.1\pm0.7$\,$M_{\rm Jup}$ \citep{Dieterich+Weinberger+Boss+etal_2018}, is difficult to reconcile with substellar cooling models at any age.  However, a subsequent dynamical analysis with much more data found a lower mass of $53.25\pm0.29$\,$M_{\rm Jup}$, fully consistent with cooling models \citep{Chen_2022}.  Two more late T dwarf companions, HD~4113~C and Gl~229~B,} 
appear much too massive to have cooled to their low observed luminosities \citep{Cheetham_2018,Brandt+Dupuy+Li+etal_2021}.  \additions{Two possible explanations for these high dynamical masses are incorrect mass measurements or faint binary companions holding a substantial amount of mass.  The alternative is a serious failure of substellar evolutionary modeling.  Gl~229~B has a $>$$100\sigma$ acceleration between Hipparcos and Gaia \citep{Brandt_2021}, making unresolved binarity an especially attractive solution despite the fact that field T dwarfs show a low binary fraction \citep{Fontanive+Biller+Bonavita+etal_2018}.} 

This paper presents conclusive evidence that Gl~229~B \subtractions{, one of the most discrepant objects with substellar evolutionary models,} is\additions{, in fact,} an unresolved binary.  Using two epochs of NIRSPEC spectroscopy on Keck, we find that Gl~229~B's RV varies from that expected from its orbit by 2.7\,km\,s$^{-1}$ and by \additions{10.4}\,km\,s$^{-1}$.  These values differ from one another and from the RV of the host star by \additions{$\approx$$11\sigma$}.  With only two epochs, we cannot derive a full orbital solution, but we can place important constraints on the properties of the system.

We find that the unseen companion, which we designate Gl~229~Bb, is at least $\approx$15\,$M_{\rm Jup}$, and could be nearly as massive as Gl~229~Ba.  The highest portion of this range is disfavored by existing Hubble Space Telescope astrometry that does not show Gl~229~B deviating from the orbit expected of its barycenter.  However, with the limited nature of these data and the range of orientations the Gl~229~B system could have, we cannot fully rule out such high companion masses. 
We find that the orbital period of the system is no more than $\approx$\additions{100}\,days and likely significantly less.  Future RV monitoring will be able to derive a precise orbit and individual masses for the two components in the Gl~229~B system. \additions{If more T dwarf companions to main sequence stars are discovered to host companions, it may point to a higher multiplicity fraction in brown dwarf companions than in the field \citep{Fontanive+Biller+Bonavita+etal_2018}.}

With at least $\approx$15\,$M_{\rm Jup}$ of the Gl~229~B system in Gl~229~Bb, the tension between Gl~229~Ba's dynamical mass and the predictions of substellar evolutionary models is mostly or entirely resolved.  Gl~229~B then stands as an example of substellar binarity, a constraint on substellar formation mechanisms, and potentially a system for which two coeval substellar objects can both have their ages, masses, spectra, and luminosities measured.  

\bibliographystyle{aasjournal}
\bibliography{refs.bib}
\label{lastpage}
\end{document}